\documentclass[conference]{IEEEtran}
\IEEEoverridecommandlockouts
\usepackage{cite}
\usepackage{amsmath,amssymb,amsfonts}
\usepackage{algorithm} 
\usepackage{algpseudocode} 
\usepackage{graphicx}
\usepackage{textcomp}
\usepackage{xcolor}
\usepackage{comment}

\def\BibTeX{{\rm B\kern-.05em{\sc i\kern-.025em b}\kern-.08em
    T\kern-.1667em\lower.7ex\hbox{E}\kern-.125emX}}
\begin{document}

\title{Comparative Analysis of Sub-band Allocation Algorithms in In-body Sub-networks Supporting XR Applications\\
\thanks{The work by Thomas Jacobsen and Ramoni Adeogun was supported by the HORIZON-JU-SNS-2022-STREAM-B-01-03 6G-SHINE project (grant agreement No. 101095738). Ramoni Adeogun's work was also partly supported by HORIZON-JU-SNS-2022-STREAM-B-01-02 project - CENTRIC (grant agreement No. 101096379)}}

\author{\IEEEauthorblockN{Saeed Bagherinejad\IEEEauthorrefmark{1},
Thomas Jacobsen\IEEEauthorrefmark{2}, 
Nuno K. Pratas\IEEEauthorrefmark{2}
and Ramoni O. Adeogun\IEEEauthorrefmark{1}}
\IEEEauthorblockA{\IEEEauthorrefmark{1}Department of Electronic Systems,
Aalborg University, 9220 Aalborg, Denmark}
\IEEEauthorblockA{\IEEEauthorrefmark{2}Nokia Standards, 9220 Aalborg, Denmark}
\IEEEauthorblockA{Email: \IEEEauthorrefmark{1}\{sbag, ra\}@es.aau.dk, \IEEEauthorrefmark{2}\{nuno.kiilerich\_pratas, thomas.jacobsen\}@nokia.com}
}

\maketitle

\begin{abstract}
In-body subnetworks (IBS) are envisioned to support reliable wireless connectivity for emerging applications including extended reality (XR) in the human body. As the deployment of in-body sub-networks is uncontrollable by nature, the dynamic radio resource allocation scheme in place becomes of the uttermost importance for the performance of the in-body sub-networks. This paper provides a comparative study on the performance of the state-of-the-art interference-aware sub-band allocation algorithms in in-body sub-networks supporting the XR applications. The study identified suitable models for characterizing in-body sub-networks which are used in a snapshot-based simulation framework to perform a comprehensive evaluation of the performance of state-of-art sub-band allocation algorithms, including greedy selection, sequential greedy selection (SG), centralized graph coloring (CGC), and sequential iterative sub-band allocation (SISA). The study shows that for XR requirements, the SISA and SG algorithms can support IBS densities up to 75\% higher than CGC.      
\end{abstract}


\section{Introduction}
With the commercialization of the fifth generation (5G) of radio technologies, the definition of the next generation has already begun. As before, the motivation behind the development of sixth-generation (6G) wireless networks is to enable the proliferation of more demanding services and technologies that previous generations, including 5G, do not fully support. Fully-immersive and multi-sensory eXtended Reality (XR) is among the main services that 6G is expected to deliver and it includes various use cases such as Augmented Reality (AR), Virtual Reality (VR), and Cloud Gaming (CG) \cite{Walid6G_Vision, Petrov2022_XR}. In these use cases, the stringent requirements on the wireless networks originate from the wireless transmission of high quality and delay sensitive video frames to the XR display device (XRDD) from external computation resources such as smartphones \cite{3GPP26928}. We envision that in-body sub-networks (IBS) could be a potential platform for supporting the XR requirements.  

IBSs are a part of the umbrella paradigm called 6G in-X sub-networks \cite{Uusitalo20216g, Berardinelli2021_InXSubn} which includes networks of specialized short-range low-power cells that serve different types of use-cases inside entities such as vehicles, robots, classrooms or the human body. Among them, IBS deals with wireless communications in the vicinity of the human body. It should be noted that, although there are existing standards such as IEEE 802.16.5 for wireless body area networks (WBAN) \cite{Movassaghi2014_WBAN}, their supported quality of service (QoS) are nowhere near the requirements of XR applications, e.g., average data rate of $30-45$Mbps, packet delay budget (PDB) of 10ms, and 99\% reliability. Supporting such demanding QoSs compared to state-of-the-art is the main motivation for developing IBS.

A key challenge for in-body sub-networks is interference, particularly in situations where a high number of people (IBSs) are assembled in a limited area, e.g., people gathering for a collective immersive social experience using XR services. There has been some works in the past on interference mitigation in 6G in-X sub-networks with a focus on in-factory sub-networks, see e.g. \cite{Adeogun2022_Enhanced, Adeogun2021_DL, SISA_2023, Adeogun2022_MAQL, Adeogun2023_MADQL, Xiao2022_MARL, Berardinelli2023_Hybrid}. In \cite{Adeogun2022_Enhanced}, a distributed and centralized dynamic sub-band group selection algorithm has been proposed for the in-X sub-networks based on the sensed interference level. A deep-learning-based algorithm has been proposed in \cite{Adeogun2021_DL} for distributed dynamic sub-band allocation which relies on offline training in a controller. In \cite{Adeogun2022_MAQL}, a multi-agent Q-learning approach is presented to joint sub-band and power allocation which relies on limited sensing information in each subnetwork. Multi-agent reinforcement learning-based algorithm in \cite{Xiao2022_MARL} utilizes received signal strength indicator (RSSI) for joint sub-band and power allocation. A centralized algorithm was proposed in \cite{SISA_2023} for sub-band allocation in sub-networks which depends on the interference to signal ratio (ISR) level reported from sub-networks. The authors of \cite{Berardinelli2023_Hybrid} introduced a hybrid framework for resource management in in-X sub-networks in which the resource allocation is done centrally or in a distributed manner based on the link state to the central controller and the battery level of the AP.

As mentioned earlier, the development of interference mitigation algorithms has been majorly focused on in-factory sub-networks. The use cases that IBS is envisioned to support, such as XR, have different data traffic patterns and also QoS levels. In addition, the signal propagation around the human body is different from other environments. These necessitate a study on the effectiveness of the resource allocation algorithms in the context of the in-body sub-networks. To enable the comparative evaluation, a methodology must be developed for modeling the IBS systems as a platform for supporting XR applications. Thus, the main contributions of this work include: i) Developing a method to model IBS networks that captures the most important aspects of IBSs and their use cases, and ii) conducting a comparative evaluation on the efficacy of different centralized and distributed sub-band allocation algorithms in IBS networks. In particular, at first, the appropriate modeling is defined for IBSs, each including a smartphone (acting as an AP and the application server for video rendering) and an XRDD. Then, the reliability of IBS networks is studied in supporting the video frame streams transmitted on the downlink (DL), from smartphones to the XRDDs. The paper is organized as follows: we start with the system model in Section~\ref{sec:SystemModel}, followed by problem formulation and a brief description of sub-band allocation algorithms in Section~\ref{sec:SAAlgorithms}. In Section~\ref{sec:PerfEval}, the numerical evaluation methodology and simulation results are presented and the conclusion is included in Section~\ref{sec:Conclusion}.

\begin{figure}[t]
    \centering
    \includegraphics[width=0.7\columnwidth]{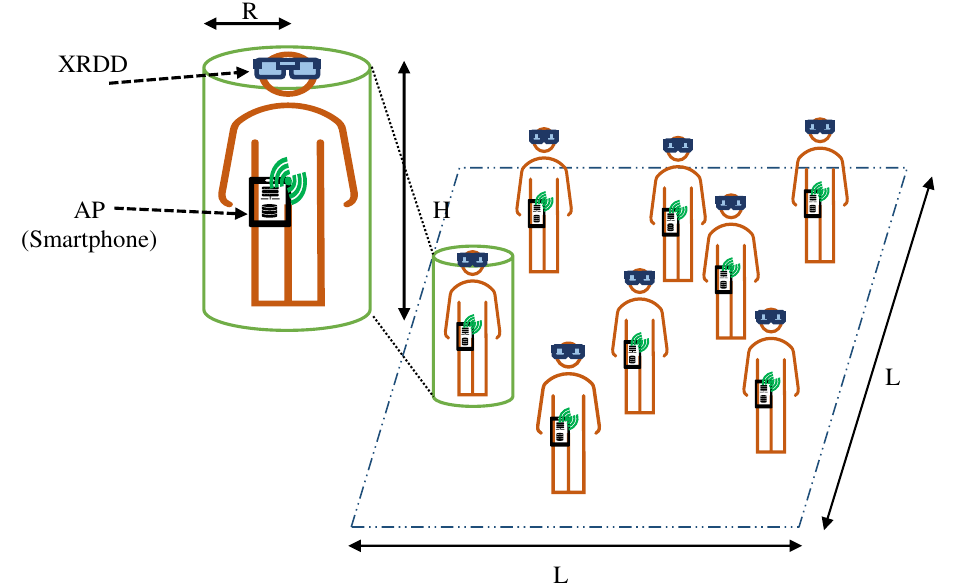}
    \caption{Deployment and shape model of $N$ In-Body sub-networks.}
    \label{fig:Deployment}
    \vspace{-10pt}
\end{figure}

\section{System Model}\label{sec:SystemModel}
\subsection{Deployment and Channel Models}
We consider a deployment comprising of $N$ in-body sub-networks (IBS) in an $L\times L$ indoor environment. Each IBS contains one smartphone acting as the AP which can coordinate with the other APs and/or the central controller. The smartphone has also the computational power to run the XR application server, as suggested in \cite{3GPP26928}. Furthermore, we assume that each IBS comprises of one head-mounted XRDD that displays the received frames. Note that in general, an IBS contains more devices such as haptic sensors, optical sensors, smartwatches, and so forth that the AP would utilize for rendering the appropriate frames. However, we focus our study on the video frame transmission in the DL from AP to XRDD which is the most important and demanding data flow of XR use cases compared to sensing feedback from other devices. That is why we only consider only an AP and an XRDD in each IBS. Furthermore, We assume that IBSs are randomly located following the Thomas Cluster Process \cite{haenggi_2012} where IBSs are located around different attraction points. These attraction points are drawn from a Homogeneous Poisson Point Process $\Phi_p$ with the density, $\lambda_p$ in the $\mathbb{R}^2$ plane. Then, the central points of IBSs are deployed as the offspring points of these attraction points where the distance from an IBS's center point to its cluster center (attraction points) follows a normal distribution 
\begin{equation}
    f_{n}(x, y|(x_{pn}, y_{pn})) = \frac{1}{\sigma_n \sqrt{2\pi}}e^{-\frac{(x-x_{pn})^2+(y-y_{pn})^2}{2\sigma_n^2}},
\end{equation} 
where the $(x_{pn}, y_{pn})$ denotes the coordinates of the attraction point and $\sigma_n$ is the standard deviation of IBSs relative distance from parent points. It is worthwhile to mention that we enforce constraints on the minimum distance between the clusters and the IBSs, i.e., $d_c^{min}$ and $d_{IBS}^{min}$, respectively. 

We consider the shape of an IBS as cylinders with a radius $R$ and height of $H$ as depicted in Figure~\ref{fig:Deployment}. Furthermore, IBSs use a dedicated spectrum with a total bandwidth of $B_t$ which is divided into $K$ equally-sized sub-bands as depicted in Figure~\ref{fig:Frame,Traffic}. The channel from the $n$-th XRDD and the $l$-th AP over the $k$-th sub-band is modeled as follows:
\begin{align}
    h_{nl}^k = \sqrt{\beta_{nl}}g_{nl}^k,
\end{align}
where $\beta_{nl}$ is the large-scale fading coefficient and $g_{nl}^k \sim \mathcal{CN}(0, 1)$ denotes the small-scale fading coefficient over the different sub-bands.

\begin{figure}[t]
    \centering
    \includegraphics[width=0.65\columnwidth]{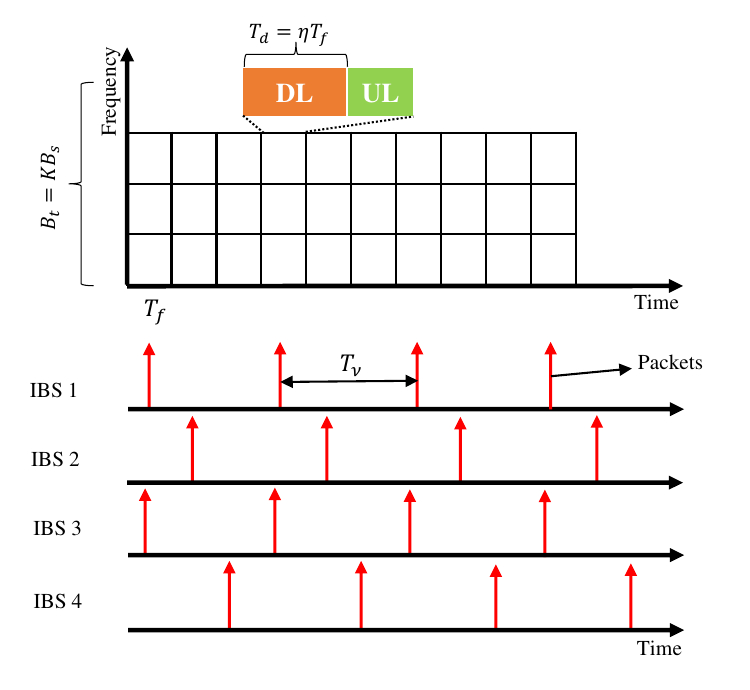}
    \caption{Frame structure and data traffic model for $T_\nu=3T_f$.}
    \label{fig:Frame,Traffic}\vspace{-10pt}
\end{figure}

\subsection{XR application and data traffic model}
The focus of our analysis is on the transmission of video frames on the DL which constitutes the most important data flow in XR applications. We have adopted the \textit{single eye buffer} model where the video frames for both eyes arrive at the same time as a single packet \cite{3GPP38838, Petrov2022_XR}. According to this periodic traffic model, packets arrive with the arrival rate equal to the frame rate, $f_\nu$ which corresponds to a packet inter-arrival time of $T_\nu =1/f_\nu$. Note that, we have considered the traffic model without jitter in the arrival of the packets. This is due to the fact that the jitter caused by changes in the routing of the packets over the network is not present for IBS networks. Moreover, the PDB determines the maximum over-the-air latency in which the packets must be transmitted successfully otherwise they are considered as dropped. The PDB for XR applications has been defined to be in the range of $5-15\text{ms}$ depending on the application, e.g., $10\text{ms}$ for AR/VR and $15$ms for CG. The packet size, $S_\nu$, is determined by the average data rate $R_\nu$ and the packet arrival rate $f_\nu$, i.e., $S_\nu = R_\nu/f_\nu$. The average data rate for the XR data source is suggested to be in the range of $R_\nu = 30-45\text{Mbps}$. Furthermore, the baseline requirement for the packet success rate is set to $99$\% which translates to a packet error rate (PER) of $10^{-2}$. Note that, this reliability constraint is with respect to the PDB.

\subsection{Transmission Frame Structure}
As depicted in Figure~\ref{fig:Frame,Traffic}, we assume a time division duplex structure where $T_f$ denotes the duration of the transmission time interval (TTI). The duration of the downlink transmission (from AP to XRDD) is denoted by $T_d=\eta T_f, 0\leq\eta\leq 1$ and the $(1-\eta)T_f$ is dedicated for uplink transmission (from sensors and XRDD to AP). We do not consider any re-transmission schemes and thus, a video frame is either transmitted successfully in a single TTI or dropped. With this assumption, the TTI duration, $T_f$, represents the over-the-air latency and therefore PDB of the packets. 

As depicted in Figure~\ref{fig:Frame,Traffic}, only a subset of all IBSs (referred to as \textit{active IBSs}) have packets to transmit at each TTI. This stems from the periodicity of the traffic model and the fact that $T_f\leq T_\nu$. To determine if the $n$-th IBS is active or not, we use a binomial random variable $b_n$. The distribution of $b_n$ depends on the packet inter-arrival time and the frame duration, i.e. $Pr\{b_n=1\}=\frac{T_f}{T_\nu}$. To prove this probability, consider a sufficiently large time interval $T_0$. In this interval, there would be $T_0/T_f$ TTIs and $T_0/T_\nu$ packets; hence, out of $T_0/T_f$ TTIs, $T_0/T_\nu$ of them would have a packet to transmit, i.e., $Pr\{b_n=1\}=\frac{T_0/T_\nu}{T_0/T_f}$. Moreover, the set of the active IBSs for a frame is defined as $\mathcal{A}=\{n|b_n=1,\; 1\leq n\leq N\}$.
We assume that this set is known for the central controller.

\section{Sub-Band Allocation Algorithms} \label{sec:SAAlgorithms}
In this section, we first describe the resource allocation problem that will be studied in this paper. Then, we will look into the sub-band allocation algorithms that were analyzed. 
\subsection{Problem description} \label{subsec:ProbDesc}
The objective of this work is to study the reliability of the IBS systems, in terms of PER, in supporting the video frame streams of XR applications in the DL. As discussed earlier, we assume that the rendering process is done on the AP of the IBS. The rendered video frames are transmitted to the XRDD for viewing by the user. The SINR expression of $n$-th active IBS ($n\in \mathcal{A}$) over $k$-th sub-band is defined as  
\begin{equation}
    \rho_n^k = \frac{a_n^k P_n |h_{nn}^k|^2}{\sum_{\substack{i\in \mathcal{A}\\ i\neq n}} a_i^k P_i |h_{ni}^k|^2 + \sigma_w^2},\; n\in \mathcal{A}
\end{equation}
where $P_n$ denotes the transmit power and $\sigma_w^2=10^{(-204+NF+10\log_{10} B_s)/10}$ is the additive white Gaussian noise (AWGN) power. Furthermore, the binary indicator $a_i^k\in \{0, 1\}$ with $\sum_{k=1}^Ka_i^k = 1$ determines if the $k$-th sub-band is allocated to $k$-th IBS or not. Using the Shannon capacity formula, the throughput of $n$-th IBS over one TTI can be described as:
\begin{equation}
    \mathcal{R}_n = \eta T_f B_s \sum_{k=1}^K a_n^k  \log_2 (1+\rho_n^k), \; n\in \mathcal{A}.
\end{equation}
Since no re-transmission is considered, a transmission would be successful if the throughput is greater than or equal to the packet size, i.e.,
\begin{equation}\label{eq:RateCondition}
    \eta T_f B_s \sum_{k=1}^K a_n^k  \log_2 (1+\rho_n^k) \geq S_\nu.
\end{equation}
This condition is used to evaluate the reliability performance of the sub-band allocation algorithms, i.e., the percentage of packets that are successfully transmitted. Using the constraint on sub-band allocation and denoting $\rho_{th} = 2^{\frac{S_\nu}{\eta T_f B_s}}$, the condition in \eqref{eq:RateCondition} translates to a condition on SINR for successful packet transmission, i.e., 
\begin{equation}\label{eq:SINR_threshold}
    \rho_n^i \geq \rho_{th},
\end{equation}
if the $i$-th sub-band is allocated to $n$-th active IBS.

\subsection{Reference Signal Transmission}
We assume that each IBS is assigned orthogonal reference signals in each sub-band in each TTI. Before starting the data transmission stage in each TTI, the active IBSs transmit their assigned reference signal with a fixed power. Then, the XRDDs measure the reference signal received powers (RSRP) on each sub-band and send them to their corresponding APs. Depending on whether the sub-band allocation algorithm is distributed or centralized, the APs either select a sub-band or report the measured RSRPs to the centralized controller. We denote the RSRP of the $n$-th XRDD from the $i$-th IBS on the $k$-th sub-band as $q_{ni}^k$, and the RSRP from the AP of the same IBS as $p_n^k$. The RSRP matrix of the $n$-th IBS is denoted by $\textbf{Q}_n\in \mathcal{R}^{K\times N}$ where

\begin{equation}
        [\textbf{Q}_n]_{ki} = 
    \begin{cases}
        q_{ni}^k, & n\neq i,\; i\in \mathcal{A},\\
        p_n^k, & n = i, \; i\in \mathcal{A},\\
        0, & i \notin \mathcal{A}.
    \end{cases} 
\end{equation}
In what follows, we will discuss how the centralized and distributed sub-band allocation algorithms utilize the available information to allocate sub-bands to each IBS.

\subsection{Centralized Sub-band Allocation}
\subsubsection{Centralized Graph Coloring (CGC)}
In this algorithm, a graph $\mathcal{G}$ is first defined by considering the sets of vertices $\mathcal{V}$ and edges $\mathcal{E}$. The vertices represent the IBSs and edges are added based on the available information. Then, using the completed graph $\mathcal{G}(\mathcal{V}, \mathcal{E})$, a vertex-coloring algorithm is used to assign a color (equivalent to a sub-band) to each vertex. The vertex-coloring assigns colors in a way so that no neighboring vertices have the same color. Thus, a vertex must be connected to the IBSs with the highest interference power so that they would be assigned different sub-bands. One way of creating the edges is using the RSRPs received from the IBSs. For this purpose, the average interference-to-signal-ratio (ISR) of the $n$-th IBS from $l$-th IBS is calculated over the $K$ sub-bands: 
\begin{equation}
    \Bar{W}_{n}(l) = \frac{1}{K}\sum_{k=1}^K\frac{q_{nl}^k}{p_n^k}.
\end{equation}
After calculating the average ISRs, the $n$-th vertex will be connected to the $K-1$ IBSs with the highest average ISRs. In addition, the averaging could be done locally on the APs which will decrease the signaling overhead of the CGC algorithm. Furthermore, the greedy vertex-coloring algorithms try to assign the minimum number of colors to the vertices. To ensure that the graph, $\mathcal{G}$, is colorable by the available number of colors (corresponding to the number of sub-bands), we propose to add the edge $(i^*, j^*)$ to the graph as follows,

\begin{equation}\label{add_edge}
    (i^*, j^*) = \underset{\underset{(i, j)\notin \mathcal{E}}{i, j \in \mathcal{A}}}{\text{arg}\,\text{max}\,\,}\Bar{W}_{i}(j),
\end{equation}
if the number of assigned colors is less than $K$. Otherwise, we remove the edge
\begin{equation}
    (i^*, j^*) = \underset{\underset{(i, j)\in \mathcal{E}}{i, j \in \mathcal{A}}}{\text{arg}\,\text{min}\,\,}\Bar{W}_{i}(j).
\end{equation}
The process of adding or removing the edges is done sequentially until the number of assigned colors equals the number of sub-bands.

\subsection{Sequential Iterative Sub-band Allocation (SISA)}
The sequential iterative sub-band allocation (SISA) \cite{SISA_2023} focuses on minimizing the sum ISR. The ISR level is calculated based on the RSRP described in the Subsection \ref{subsec:ProbDesc} as follows:
\begin{equation}
    W_n(l, k) = \frac{q_{nl}^k}{p_n^k}, \; 1\leq l\leq N.
\end{equation}
This measurement matrix of size $N\times K$ is calculated at the AP and then forwarded to the central controller. After receiving the ISRs from all IBSs, the controller runs the sequential iterative sub-band allocation algorithm which is described in \textbf{Algorithm \ref{alg:SISA}}. This algorithm starts first with a random allocation that we denote by the mapping function $\mathcal{C}^0(n), \forall n\in \mathcal{A}$. Moreover, $\mathcal{B}^0_k = \{n|n\in \mathcal{A}, \mathcal{C}^0(n)=k\}$ denotes the set of active users with the initial allocated sub-band of $k$. After initialization, the algorithm goes through the active IBSs sequentially and chooses the sub-band with the lowest mutual ISR according to the current allocation which is described in line 5 of the algorithm. After selecting the sub-band, the allocation is updated at line 6 before moving to the next IBS. This process is carried out for $M$ iterations.

\begin{algorithm}[!t]
	\caption{Sequential Iterative Sub-band Allocation}\label{alg:SISA}
	\begin{algorithmic}[1]
		\Require $ W_n(l, k), n, l=1,\ldots,N,\; k=1,\ldots,K$\\
            \textbf{Initialize:} $\mathcal{C}^0(n), \forall n\in\mathcal{A}$, $\mathcal{B}^0_k = \{n|n\in \mathcal{A}, \mathcal{C}^0(n)=k\}$
			\For {$m=1,2,\ldots,M$}
                    \For {$n=1,2,\ldots,|A|$}
				    \State $t=|A|(m-1)+n$ 
                        \State $\mathcal{C}^t(n) = \underset{1\leq k\leq K}{\text{arg}\,\text{min}}\,\, \sum_{\underset{l\neq n}{l\in \mathcal{B}^{t-1}_k}} W_n(l, k) + W_l(n, k)$ 
                        \State $\mathcal{B}^t_k = \{n|n\in \mathcal{A}, \mathcal{C}^t(n)=k\}$
                    \EndFor
			\EndFor
		\Ensure $\mathcal{B}^d_k$, where $d=|A|\times M$
	\end{algorithmic} 
\end{algorithm}
\subsection{Distributed Sub-band Allocation}
\subsubsection{Greedy Selection}
Greedy selection is the simplest distributed algorithm where IBSs after receiving reference signals, independently select a $k^*$ as follows:
\begin{equation}
    k^* =  \underset{k=1,...,K}{\text{arg}\,\text{min}\,\,} \sum_{l\in \mathcal{A}} \frac{q_{nl}^k}{p_n^k}.
\end{equation}
This algorithm would significantly reduce the signaling overhead. The overhead can be further decreased by removing the reference signal transmission. Then, the IBSs have to rely on the sensed aggregate interference level during the data transmission stage.  

\subsubsection{Sequential Greedy Selection}
In this algorithm, there is a setup stage before data transmission during which the active IBSs coordinate to select sub-bands. After the RS transmission phase and based on a predefined order, the first IBS selects a sub-band and broadcasts the selection to the other IBSs. Each subnetwork stores a local copy of the sub-band selection sets $B_k$, denoting the set of active IBSs that have selected the $k$-th sub-band, and updates them according to the broadcast on sub-band selection. Then, the next IBS will choose a sub-band based on the information it has received about the sub-band selection and the ISR levels. The $n$-th IBS will choose the $k^*$ based on the following criteria:
\begin{equation}
    k^*=\underset{k=1,...,K}{\text{arg}\,\text{min}\,\,} \sum_{l\in \mathcal{B}_k} \frac{q_{nl}^k}{p_n^k},
\end{equation}
This process is carried out $M$ times which is set beforehand. One important note here is that since IBSs take turns selecting a sub-band and informing others, the algorithm might not be practically feasible to be carried out in a limited preamble time before data transmission, especially in dense scenarios. 

Finally, we also evaluate the performance of random sub-band selection as the worst-case baseline for the sub-band allocation algorithms. With this scheme, the IBSs randomly select one of the sub-bands to transmit their data.  

\section{Performance Evaluation}\label{sec:PerfEval}
\subsection{Simulation Assumptions and Methodology}
We have considered a $20\text{m}\times20\text{m}$ environment where there are $5$ attraction points as the parent points of the Thomas Cluster Process that corresponds to $\lambda_p=12.5\times10^3\,\text{points}/\text{km}^2$. After deploying these parent points, the IBSs are dropped according to a normal distribution with $\sigma_n = 2\text{m}$. The total number of IBSs is considered to be $\{30, 40,\ldots, 100\}$ which is equivalent to $\lambda$ from $7.5\times10^4 \text{IBS}/\text{km}^2$ to $25\times10^4\text{IBS}/\text{km}^2$. The shape of the IBSs is assumed to be a cylinder with a radius $R=0.25$m and a height of $H=1.9$m. To capture the difference in the heights of people, the z-coordinate of the APs and XRDDs are generated to be uniformly distributed between $0.9\text{m} - 1.15\text{m}$ and $1.65\text{m} - 1.90\text{m}$, respectively. 

Furthermore, a bandwidth of $B_t=1\text{GHz}$ with the carrier frequency of $f_c=2.4\text{GHz}$ is considered which is divided into $K=\{5, 10, 15,\ldots, 30\}$ sub-bands. The TTI duration is assumed to be $T_f=5\text{ms or }10\text{ms}$ to evaluate the effect of it on the reliability. A summary of the simulation parameters is presented in Table~\ref{tab:SimValue}. The large scale fading for intra-body channel is modeled as $\beta_{nn}(dB) = 8.6\log_{10}(d_{nn}) + 46.1+ 2\chi$ where the log-normal shadowing factor is denoted by $\chi \sim \mathcal{N}(0,1)$ \cite{Takizawa_CM}. The 3-D distance between the $n$-th XRDD and the corresponding AP is denoted by $d_{nn}$ in meters. The large-scale fading for the inter-body channel is assumed to follow the proposed model for D2D communications by 3GPP \cite{3GPP36843},

\begin{align}
    \beta_{nl}(dB) = 
    \begin{cases}
        38.8 + 16.9\log_{10}(d_{nl}) + 3 \chi, & S_{nl}=LOS\\
        17.5 + 43.3\log_{10}(d_{nl}) + 4 \chi, & S_{nl}=NLOS,
    \end{cases} 
\end{align} 
for $n\neq l$. $d_{nl}$ denotes the 3-D distance of $n$-th XRDD and the $l$-th AP in meters. The variable $S_{nl} \in\{LOS, NLOS\}$ determines the state of the inter-body link. We assume that $S_{nl}$ follows the probability distribution
\begin{equation}
    Pr\{S_{nl}=LOS\} =\exp(-\gamma d_{nl}^{2D}),
\end{equation}
where $\gamma=2\lambda R$ and $\lambda$ is the deployment density of IBSs in the environment. Furthermore, $d_{nl}^{2D}$ is the 2-D distance of transceivers in the xy-plane. This model is based on the consideration of the human body (IBSs) with the shape of a cylinder as the blockages. The results in the following section are calculated based on $5\times10^6$ snapshots. 

\begin{table}[t] 
\centering
\caption{Values of Simulation Parameters}
\label{tab:SimValue}
\resizebox{0.8\columnwidth}{!}{
\begin{tabular}{|ll|} 
\hline
Parameter                              & Value \\ \hline
Deployment Area ($\text{m}^2$) & $20\text{m}\times 20\text{m}$ \\
Number of IBSs & $\{30, 40,\ldots, 100\}$ \\
Number of Clusters & $5$, ($\lambda_p=12.5\times 10^3/\text{km}^2$) \\ 
Std. of IBSs Relative Location, $\sigma_n$ (m) & $2$ \\
Minimum Distance of Clusters and IBSs (m) & $d_c^{min}=4$, $d_{IBS}^{min}=0.5$\\
IBS Cylinder Dimensions ($\text{m}$)      & $R=0.25$, $H=1.9$\\
Total Bandwidth, $B_t$ (GHz)          & $1$ \\
Carrier Frequency, $f_c$ (GHz) & $2.4$ \\
Number of Sub-bands, $K$            & $\{5, 10, 15,\ldots, 30\}$ \\
Transmission Frame Duration, $T_f$ (ms) & $5 \text{ or } 10$\\ 
Percentage of Downlink Transmission, $\eta$ & 0.8\\
Average Data Rate, $R_\nu$ (Mbps) & $45$\\
Packet Inter-arrival Time, $T_\nu$ (ms)& $16.67$ \\
Transmit Power (dBm) & $0$\\
Noise Figure (dB) & 10 \\
\hline
\end{tabular}
}
\vspace{-10pt}
\end{table}

\subsection{Simulation Results}
\begin{figure}[t!]
    \centering
    \includegraphics[width=0.6\columnwidth]{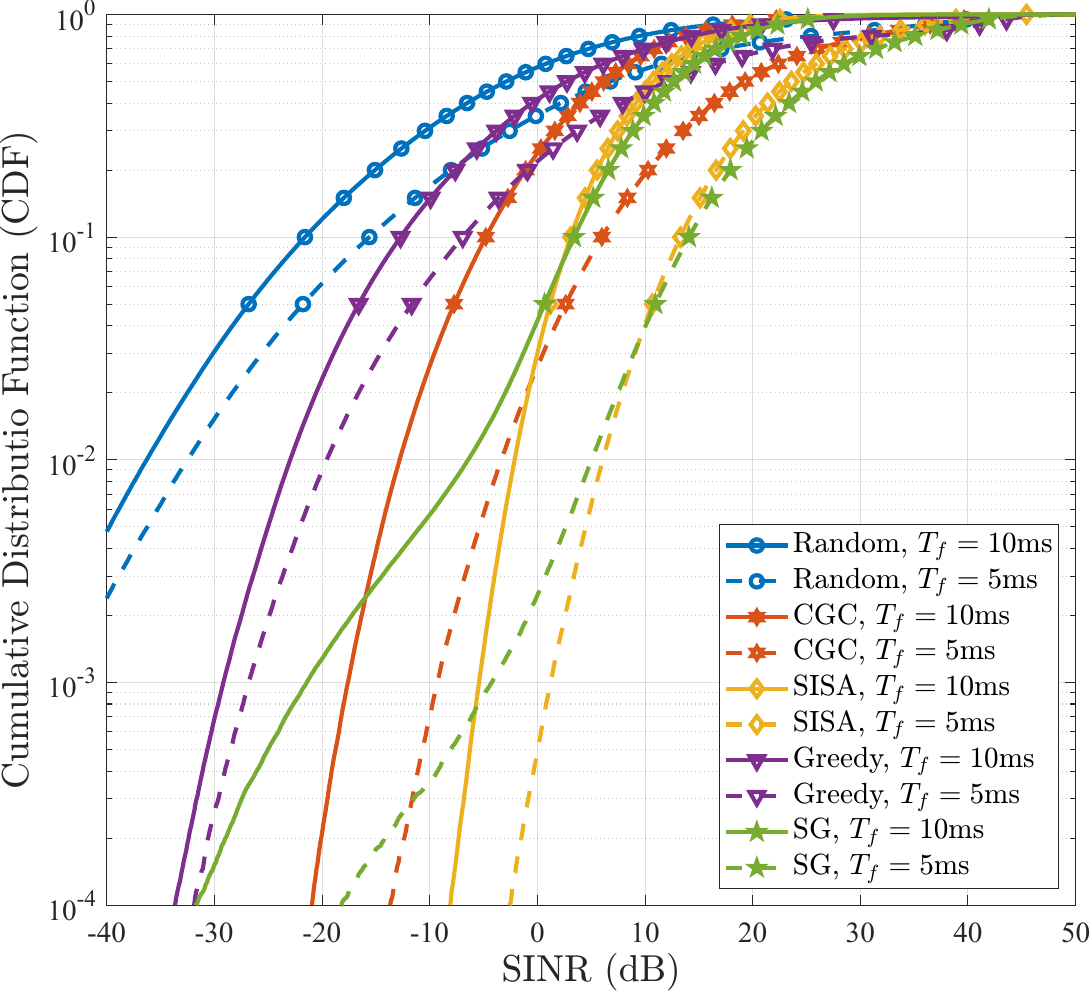}
    \caption{CDF of SINR, for $N=60$, $K=10$, $T_f=5,10$ms.}
    \label{fig:CDFofSINR5ms10ms}\vspace{-10pt}
\end{figure}
\begin{figure}
    \centering
    \includegraphics[width=0.6\columnwidth]{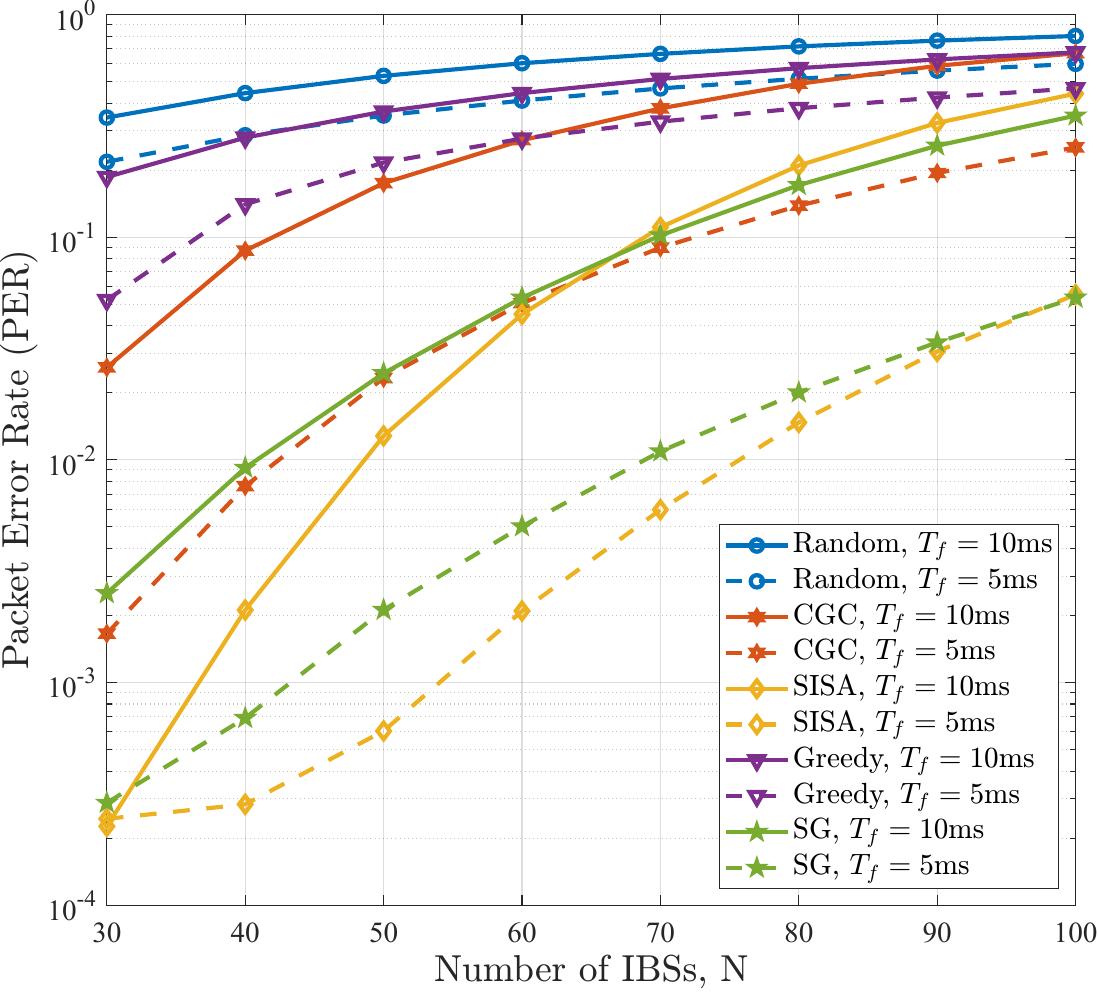}
    \caption{Packet Error Rate (PER) versus the Number of IBSs ($N$), for $K=10$ and $T_f=5\text{ms}, 10$ms.}
    \label{fig:PER_vs_N_T5ms10ms}\vspace{-10pt}
\end{figure}
\begin{figure}[t]
    \centering
    \includegraphics[width=0.6\columnwidth]{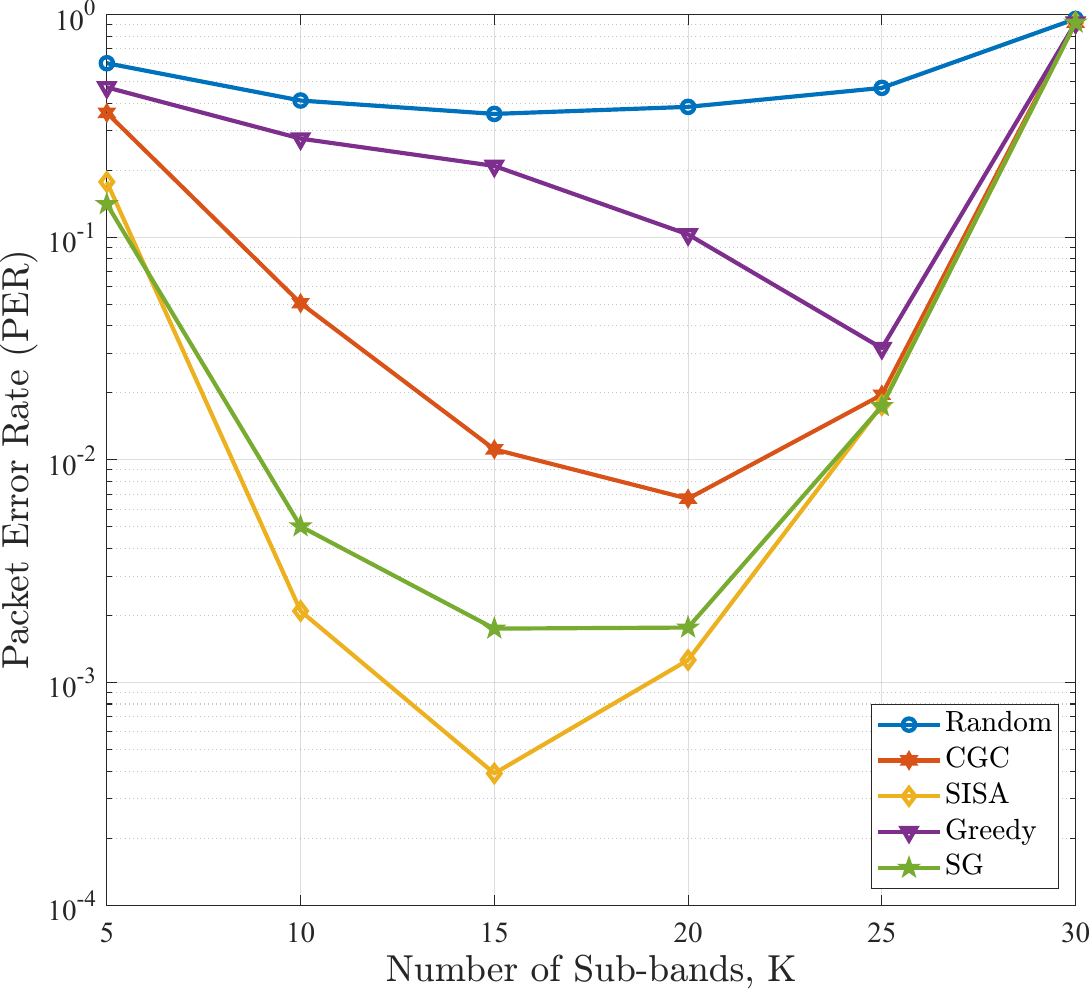}
    \caption{Packet Error Rate (PER) versus the Number of Sub-bands ($K$), for $N=60$ and $T_f=5$ms.}
    \label{fig:PER_vs_K_T5}\vspace{-10pt}
\end{figure}
Figure~\ref{fig:CDFofSINR5ms10ms} depicts the cumulative distribution function (CDF) of SINR for the sub-band allocation algorithms described in Section~\ref{sec:SAAlgorithms} for $T_f=5\text{ms and } 10\text{ms}$. As seen in this figure, the worst performing sub-band allocation algorithm, Greedy, shows more than $10$dB gain compared to random selection in the tail of the CDF. SISA has at least $10\text{dB}$ higher SINR in the tail compared to the other algorithms. SG outperforms the SISA in the high percentiles, but, it fails in the tail which is likely due to the impact of the mutual ISR that is not considered. In addition, it is shown that lower TTI duration yields higher SINR due to a lower number of active IBSs per TTI which is evident from the SINR increase in $T_f=5$ms compared to $T_f=10$ms.

Furthermore, Figure~\ref{fig:PER_vs_N_T5ms10ms} shows the packet error rate (PER) versus the number of IBSs. The PER is evaluated for two different $T_f$ values. The figure shows that SISA and SG are the best-performing sub-band allocation algorithms due to their superiority in mitigating interference. By considering the baseline PER of $10^{-2}$ as suggested in \cite{3GPP26928}, SISA and SG can support more than around $N=70$ IBSs for $T_f$ is $5$ms. This is around 75\% more than CGC. Second, although higher $T_f$ decreases the threshold on SINR according to \eqref{eq:SINR_threshold}, it also reduces the SINR by increasing the number of active IBSs per TTI, as seen in Figure~\ref{fig:CDFofSINR5ms10ms}. The effect of a larger number of active IBSs overcomes the advantage of having a lower SINR threshold which is seen in Figure~\ref{fig:PER_vs_N_T5ms10ms} with higher PER for $T_f=10$ms compared to $T_f=5$ms.  

Figure~\ref{fig:PER_vs_K_T5} illustrates the impact of $K$, number of sub-bands, on PER for $N=60$ and $T_f=5\text{ms}$. It is shown that for all sub-band allocation algorithms, PER decreases at first by increasing the $K$ which stems from the higher degree of interference diversity, due to the higher number of sub-bands. However, after a certain number of sub-bands, e.g., $K=15$ for SISA, the PER begins to increase with the $K$. This is an indication that having more narrower sub-bands is no longer beneficial. This is reasonable since a higher number of sub-bands with fixed total bandwidth means smaller sub-band bandwidth for data transmission and therefore, putting a more strict constraint for successful packet transmission based on \eqref{eq:SINR_threshold}. This effect is evident in a rapid increase of PER in Figure~\ref{fig:PER_vs_K_T5} for all sub-band allocation algorithms. 

\subsection{Signaling Overhead of the sub-band allocation algorithms}
Apart from the performance in mitigating interference, another aspect that differentiates the sub-band allocation algorithms is the sensing and signaling overhead that they incur. There are two sources for the signaling overhead of the sub-band allocation algorithms: 1) transmission of the reference signals, and 2) feedback of the measurements. Transmitting the reference signals has an overhead in the order of $\mathcal{O}(NK)$ since the systems need $N$ orthogonal sequences that IBSs have to transmit over $K$ sub-band which is the same for all the algorithms. CGC depends on the ISR levels reported by the sub-networks to create the graph and assign the colors. Each IBS has to report a vector of size $N$ to the centralized controller. Thus, the signaling overhead of CGC is in order of $\mathcal{O}(N^2)$. Since the number of sub-bands, $K$ is typically much less than the number of IBSs, the overhead incurred by CGC is dominated by $\mathcal{O}(N^2)$. 

For SISA, each IBS must report individual ISR levels on all the $K$ sub-bands translating to signalling of a matrix of size $N\times K$. Overall, the asymptotic signaling overhead of the SISA will be in the order of $\mathcal{O}(N^2)$. For the centralized algorithms, there is also an overhead of reporting the allocation to the IBSs which is in order of $N$ and is negligible compared to other parts of signaling overhead. Furthermore, the greedy algorithm only requires the devices to report the sensed aggregate ISR levels over sub-bands to their APs which results in the overhead in order of $\mathcal{O}(N)$. The SG algorithm has a signaling overhead in the order of $\mathcal{O}(NM)$ as each IBS must broadcast its sub-band selection $M$ times to the whole network. In conclusion, SISA and CGC have higher overhead than SG and greedy which is caused majorly by the report of measurements to the central controller. 

\section{Conclusions} \label{sec:Conclusion}
We considered the problem of sub-band allocation for IBS supporting the XR downlink video frames stream. We evaluated the performance of state-of-the-art sub-band allocation methods and their capability to mitigate interference and showed that these algorithms could help improve IBS network performance compared to non-interference-aware algorithms. It is illustrated that the SISA and SG algorithms have the best performance in terms of PER; however, SISA incurs higher signaling overhead to the network which results in poor scalability. In addition, the results show that a higher number of sub-bands and larger TTI duration will not necessarily result in a better packet success rate. Finally, we can conclude that although IBS can act as a platform for XR applications, further improvements are needed in resource allocation algorithms for supporting ultra-dense deployments. 

\bibliographystyle{ieeetr}
\bibliography{refIBS}
\end{document}